\documentclass[prl,twocolumn,showpacs,amsmath,amssymb]{revtex4}
\usepackage{graphicx}
\usepackage{dcolumn}
\usepackage{bm}
\begin{document}
\title{Effect of Two Level System Saturation on 
Charge Noise in Josephson Junction Qubits} 
\author{Clare C. Yu$^1$, Magdalena Constantin$^1$, and John M. Martinis$^2$}
\affiliation{
$^1$ Department of Physics and Astronomy, University of California, 
Irvine, California 92697\\
$^2$ Department of Physics, University of California, Santa Barbara, 
California 93106}
\date{\today}
\pacs{74.40.+k,85.25.Cp,77.22.Gm,03.67.Pp}
\begin{abstract}
We show that charge noise $S_Q$ in Josephson qubits can be produced by 
fluctuating two level systems (TLS) with electric dipole moments in the 
substrate using a flat density of states. At high frequencies the frequency 
and temperature dependence of the charge noise depends on the ratio 
$J/J_c$ of the electromagnetic flux $J$ to the critical flux $J_c$. 
It is not widely appreciated that TLS in small qubits can easily 
be strongly saturated with $J/J_c\gg 1$. Our results are consistent 
with experimental conclusions that $S_Q\sim 1/f$ at low frequencies 
and $S_Q\sim f$ at high frequencies. 
\end{abstract}

\pacs{74.40.+k, 03.65.Yz, 03.67.-a, 85.25.-j}
\maketitle

Noise and decoherence are a major obstacle to using superconducting
Josephson junction qubits to construct quantum computers. Recent
experiments \cite{Simmonds2004,Martinis2005} indicate that a dominant
source of decoherence is two level systems (TLS) in the insulating
barrier of the tunnel junction as well as in the dielectric material
used to fabricate the circuit. It is believed that these TLS fluctuators
lead to low frequency $1/f$ charge noise $S_Q(f)$
\cite{Martinis1992,Mooij1995,Zorin1996,Kenyon2000,Astafiev2006}.
However, at high frequencies, one experiment finds that
the charge noise increases linearly with 
frequency \cite{Astafiev2004}. This has prompted some theorists to 
use a TLS density of states linear in energy \cite{Shnirman2005} which is
contrary to the constant density of states that has been so successful
in explaining the low temperature properties of glasses such as
the specific heat that is linear in temperature \cite{Phillips}.
A linear distribution has been proposed in conjunction with a Cooper 
pair tunneling into a pair of electron traps \cite{Faoro2005}, and
with electron hopping between Kondo--like traps to account
for the charge noise \cite{Faoro2006}. 

However, these previous theoretical efforts have neglected the important 
issue of the saturation of the two level systems. Dielectric (ultrasonic) 
experiments on insulating glasses at low temperatures have found that 
when the electromagnetic (acoustic) energy flux $J$ used to make the measurements 
exceeds the critical flux $J_c$, the dielectric (ultrasonic) power 
absorption by the TLS is saturated, and the attenuation decreases
\cite{Golding1976,Arnold1976,Schickfus1977,Graebner1983,Martinis2005}. 
The previous theoretical efforts to explain the linear increase 
in the charge noise in Josephson junctions assumed 
that the TLS were not saturated, i.e., that $J\ll J_c$. 
This seems sensible since the charge noise experiments were done 
in the limit where the qubit absorbed only one photon. However,
stray electric fields could saturate TLS in the dielectric substrate
as the following simple estimate shows. We can estimate the voltage $V$ 
across the capacitor associated with the substrate and ground plane
beneath the Cooper
pair box by setting $CV^2/2=\hbar\omega$ where $\hbar\omega$ is the
energy of the microwave photon. We estimate the capacitance 
$C=\varepsilon_o\varepsilon A/L \sim 7$ aF using the area $A=40\times 800$ nm$^2$
\cite{Astafiev2004} of the Cooper pair box, the thickness $L=400$ nm of the
substrate \cite{Astafiev2004}, and the dielectric constant $\varepsilon=10$. 
Using $\omega/2\pi=10$ 
GHz, we obtain a voltage of $V\sim 1.4\;$mV. The substrate 
thickness $L$ of 400 nm yields an electric field of 
$E\sim 3.4\times 10^{3}$ V/m. For amorphous SiO$_2$ at 
$f=7.2$ GHz and SiN$_{x}$ at $f=4.7$ GHz, the critical rms 
voltage $V_{c} \sim 0.2\;\mu$V \cite{Martinis2005}, and with 
a capacitor thickness of 300 nm, the critical field is 
$E_c\sim 0.7$ V/m at $T=25$ mK. So $E/E_{c}\sim 5\times 10^{3}$, 
and $J/J_{c}=\left(E/E_{c}\right)^{2}\sim 2\times 10^{7}\gg 1$. 
A similar estimate shows that a single photon would
even more strongly saturate resonant TLS in the insulating barrier
of the tunnel junction, again resulting in $J/J_{c}\gg 1$.
However, there are only a few TLS in the oxide barrier of a 
small tunnel junction. (For a parallel plate capacitor with a specific 
capacitance of 60 fF/$\mu$m$^2$, $L=1.5$ nm, $A=1\;\mu$m$^2$ and a 
dielectric constant of 10, the volume is $\Omega=1.5\times 10^{-21}$ m$^3$.
With a density of states 
$P_0\simeq 10^{45}/\left(Jm^{3}\right)\simeq663/h$GHz$\mu$m$^3$,
there are only 2 TLS with an energy splitting
less than 10 GHz.) A single fluctuator would have a Lorentian
noise spectrum. The presence of $1/f$ noise implies many more
than 2 fluctuators. It is likely that
these additional fluctuators are in the substrate.
Our main point is that TLS in small devices are easily saturated. 

In this letter we explore the consequences of this saturation. 
We find that at high frequencies ($\hbar\omega\gg k_B T$), 
the frequency and temperature dependence of the charge 
noise depends on the ratio $J/J_c(\omega,T)$ of the 
electromagnetic flux $J$ to the critical value $J_c(\omega,T)$ 
which is a function of frequency $\omega$ and temperature $T$. 
Starting from the fluctuation-dissipation theorem, we show that 
the charge noise is proportional to the dielectric loss 
tangent $\tan\delta$. We then calculate the dielectric loss tangent due to 
fluctuating TLS with electric dipole moments 
\cite{Phillips,Classen1994,Arnold1976}. At low frequencies we 
recover $1/f$ noise. At high frequencies $\tan\delta$ 
is proportional to $1/\sqrt{1+\left(J/J_c(\omega,T)\right)}$. 
In the saturation regime ($J\gg J_c(\omega,T)$), $\tan\delta$ 
and hence, the charge noise, are proportional to 
$\sqrt{J_c(\omega,T)/J}$. Some TLS experiments 
\cite{Bachellerie1977,Bernard1979,Graebner1979} 
indicate that $J_c(\omega,T)\sim\omega^{2}T^{2}$ which implies that 
at high frequencies the charge noise and the dielectric loss tangent 
would increase linearly in frequency if $J\gg J_c(\omega,T)$. 
Unlike previous theoretical efforts, we use the standard TLS density 
of states that is independent of energy, and can still 
obtain charge noise that increases linearly with frequency in agreement
with the conclusions of Astafiev {\it et al.} \cite{Astafiev2004}.

In applying the standard model of two level systems to Josephson 
junction devices, we consider a TLS that sits in the insulating substrate 
or in the tunnel barrier,
and has an electric dipole moment ${\bf p}$ consisting
of a pair of opposite charges separated by a distance $d$. 
The electrodes are located at $z=0$ and $z=L$ 
and kept at the same potential. The angle between ${\bf p}$ and 
$z$--axis which lies perpendicular to the plane of the electrodes
is $\theta_0$. The dipole flips and induces charge fluctuations 
on the electrodes. These induced charges are 
proportional to the $z$-component of the dipole moment, i.e., 
$Q=|p\cos \theta_0/L|$. 

The TLS is in a double--well potential with a tunneling matrix 
element $\Delta_0$ and an asymmetry energy $\Delta$ 
\cite{Phillips}. The Hamiltonian of a TLS 
in an external ac field can be written as $H=H_0+H_1$, 
where $H_0=\frac{1}{2}(\Delta \sigma_z+\Delta_0 \sigma_x)$, 
and $H_1=-\sigma_z {\bf p}\cdot\bm{\xi}_{ac}(t)$. Here 
$\sigma_{x,z}$ are the Pauli spin matrices and 
$\bm {\xi}_{ac}(t)=\bm {\xi}_{ac}\mbox{cos}\omega t$ is a small 
perturbing ac field of frequency $\omega$ that couples to the 
TLS electric dipole moment. After diagonalization, the TLS 
Hamiltonian becomes $H_0=\frac{1}{2}E\sigma_z$, where $E$ is 
the TLS energy splitting (i.e., $E=\sqrt{\Delta^2 + \Delta_{o}^2}$) 
and, in the new basis, the perturbing Hamiltonian is 
$H_1=-(\sigma_z\Delta/E+\sigma_x\Delta_0/E){\bf p}\cdot\bm{\xi}_{ac}(t)$. 
The complete TLS Hamiltonian is similar to the Hamiltonian 
$H_S = -\gamma{\bf S}\cdot{\bf B}$ for a spin $1/2$ particle 
in a magnetic field given by ${\bf B}(t)={\bf B}_0+{\bf B}_1(t)$, 
where the static field is $-\hbar \gamma {\bf B}_0=(0,0,E)$ 
and the rotating field is 
$\hbar \gamma {\bf B}_1(t)=(2\Delta_0/E,0,2\Delta/E)
{\bf p}\cdot \bm {\xi}_{ac}\mbox{cos}\omega t$ \cite{herve}. 
$\gamma$ is the gyromagnetic ratio 
and ${\bf S}=\hbar\bm {\sigma}/2$. Therefore, the equations 
of motion of the expectation values of the spin components 
are given by the Bloch equations 
\cite{Slichter} with the longitudinal and the transverse 
relaxation times, $T_1$ and $T_2$. In the standard model, an excited 
two-level system decays to the ground state by emitting a phonon. 
The longitudinal relaxation rate, $T_1^{-1}$, is given by 
\cite{Phillips} 
\begin{equation}
T_1^{-1}=a
E \Delta_o^{2}\coth\left(\frac{E}{2k_BT}\right), 
\label{eq:T1inv}
\end{equation}
where $a=\left[\gamma_d^2/\left(2\pi\rho\hbar^4\right)\right]
\left[\left(1/c_{\ell}^{5}\right)+\left(2/c_{t}^{5}\right)\right]$
where $\rho$ is the mass density, $c_{\ell}$ is the longitudinal 
speed of sound, $c_{t}$ is the transverse speed of sound, and
$\gamma_d$ is the deformation potential. 
The distribution of TLS parameters can be expressed in terms of 
$E$ and $T_1$: $P(E,T_1)=P_0/(2T_1\sqrt{1-\tau_{min}(E)/T_1})$ 
\cite{Phillips1987,Phillips} where $P_0$ is a constant. The minimum 
relaxation time $\tau_{min}(E)$ corresponds to a symmetric 
double--well potential (i.e., $E=\Delta_0$). The transverse 
relaxation time $T_2$ represents the broadening of levels due 
to their mutual interaction \cite{Black1977}.

{\it General Expression for Charge Noise:}
We derive a general expression valid at all frequencies
for the charge noise in terms of the dielectric loss tangent 
$\tan\delta(\omega)=\varepsilon^{\prime\prime}(\omega)/
\varepsilon^{\prime}(\omega)$, 
where $\varepsilon^{\prime}(\omega)$ and 
$\varepsilon^{\prime\prime}(\omega)$ are the real and imaginary parts of 
the permittivity. 

According to the Wiener-Khintchine theorem, the charge spectral density 
$S_Q(\omega)$ is twice the Fourier transform $\Psi_Q(\omega)$ 
of the autocorrelation function of the fluctuations in the charge.
From the fluctuation-dissipation theorem, the charge noise is given by:
\begin{equation}
S_{Q}(\bm{k},\omega)=\frac{4\hbar}{1-\mbox{e}^{-\hbar \omega/k_BT}}
\chi_{Q}^{\prime\prime}(\bm{k},\omega),
\label{eq:S_Q}
\end{equation}
where $Q$ is the induced (bound) charge and 
$\chi_{Q}^{\prime\prime}(\bm{k},\omega)$ is the Fourier transform of
$\chi_{Q}^{\prime\prime}(\bm{r},t;\bm{r}^{\prime},t^{\prime})=
\langle\left[Q(\bm{r},t),Q(\bm{r}^{\prime},t^{\prime})\right]\rangle/2\hbar$. 
We use $Q=\int \bm{P}\cdot d\bm{A}$, where $\bm{P}$ is the electric 
polarization density, and choose 
$P_z$ and $d\bm{A} \| \hat{z}$ since $Q\sim |p_z|$ to find
$\chi_{Q}^{\prime\prime}(\bm{k},\omega)=\varepsilon_{o}A^2
\chi_{P_z}^{\prime\prime}(\bm{k},\omega)$, where $\varepsilon_0$ is 
the vacuum permittivity, $A$ is the area of a 
plate of the parallel plate capacitor with capacitance $C$, and 
$\chi_{P_z}^{\prime\prime}(\bm{k},\omega)$ is the imaginary part 
of the electric susceptibility. Setting $\bm{k}=0$, and using
$\varepsilon_o\chi_{P_z}^{\prime\prime}(\omega)=
\varepsilon^{\prime}(\omega)\tan\delta(\omega)$, and
$C=\varepsilon^{\prime}A/L$, we find
\begin{equation}
S_Q(\omega)=\frac{4\hbar C}{1-\mbox{e}^{-\hbar \omega/k_BT}} 
\tan\delta(\omega),
\label{eq:S_Q_Tan_delta}
\end{equation}
where $S_Q(\omega)\equiv S_Q(\bm{k}=0,\omega)/\Omega$, the volume of 
the capacitor is $\Omega=AL$, and $\varepsilon^{\prime}(\omega)=
\varepsilon^{\prime}+\varepsilon_{TLS}(\omega)\simeq\varepsilon^{\prime}=
\varepsilon_0 \varepsilon_r$ where $\varepsilon_r$ is the relative permittivity.
The frequency dependent $\varepsilon_{TLS}(\omega)$ produced by TLS is negligible 
compared to the constant permittivity $\varepsilon^{\prime}$ \cite{Phillips}. 

The dynamic electric susceptibilities ($\chi^{\prime}(\omega)$,
$\chi^{\prime\prime}(\omega)$), and hence the dielectric loss 
tangent, can be obtained by solving the Bloch 
equations \cite{Jackle1975,Arnold1976,Graebner1983}. 
(Shnirman {\it et al.} \cite{Shnirman2005} gave a 
reformulation of the Bloch equations for TLS in terms of 
density matrices and the Bloch-Redfield theory.) 
The electromagnetic dispersion and attenuation due to TLS has two 
contributions. First there is the relaxation process due to 
the modulation of the TLS energy splitting by the incident wave 
resulting in readjustment of the equilibrium level population. 
This is described by $\chi_{z}(\omega)$ that comes 
from solving the Bloch equations for $S_z$ 
which is associated with the population difference of 
the two levels. The second process is the resonant absorption 
by TLS of phonons with $\hbar \omega=E$. Resonance is associated 
with $\chi_{\pm}(\omega)=\chi_{x}(\omega) \pm i\chi_{y}(\omega)$ 
since $\sigma_x$ and $\sigma_y$ are
associated with transitions between the two levels. The total 
dielectric loss tangent is the sum of these two contributions: 
$\tan \delta=\tan \delta_{REL} +\tan \delta_{RES}$. 
The steady-state solution of the Bloch equations and the resulting 
dielectric loss tangent for TLS are well known 
\cite{Jackle1975,Arnold1976,Graebner1983,herve}. 

In the steady-state 
regime the experimental values for the relaxation times $T_1$ and $T_2$ 
are considered small compared to the electromagnetic pulse duration, $t_p$. 
One might ask if one should use the transient solution
\cite{herve} of the Bloch equations since the pulse applied to 
superconducting qubits is often extremely short, 
$t_p\sim10^{-10}$ sec \cite{Astafiev2004}. We find that
the transient $z$--component of the magnetization
\cite{herve}, $S_z^{0}(t)$, decays exponentially to the 
equilibrium value denoted by $S_{z,eq}^0$, i.e.: 
$S_z^{0}(t)=S_{z,eq}^0+\mbox{exp}(-t/T^{\star})[S_z^{0}(0)-S_{z,eq}^0]$, 
where $S_z^{0}(0)$ is the initial value of $S_z^{0}$ and 
$S_{z,eq}^0=-\hbar\mbox{tanh}(E/2k_BT)/2$. 
The transient relaxation time $T^{\star}$ is given by
\begin{align}
&T^{\star}=\frac{T_1}{1+\left(J/J_c(\omega,T)\right) 
\times g(\omega,\omega_0,T_2)},~~~ \mbox{where} \nonumber\\
&g(\omega,\omega_0,T_2)=\frac{1}{1+T_2^2(\omega-\omega_0)^2}+
\frac{1}{1+T_2^2(\omega+\omega_0)^2}.
\end{align}
Here $J/J_c(\omega,T)=({\bf p^{\prime}}\cdot \bm {\xi}_{ac}
/\hbar)^2 T_1 T_2$, ${\bf p^{\prime}}=(\Delta_0/\epsilon){\bf p}$ 
represents the  induced TLS dipole moment, and 
$\omega_0=\gamma B_0=-E/\hbar$. In 
the saturated regime, for $J\gg J_c(\omega,T)$, we find that  
$T^{\star}\approx [T_2({\bf p^{\prime}}\cdot \bm {\xi}_{ac}/\hbar)^2]^{-1}$. 
Using $p^{\prime}=3.7$ D, $\xi_{ac}=3.4\times 10^3$ V/m, 
and $T_2=8$ $\mu$s at $T=0.1$ K \cite{Bernard1979,herve} 
yields $T^{\star}\simeq 8\times10^{-13}$ sec at resonance 
when $\omega/2\pi\simeq \omega_0/2\pi=10$ GHz. This value 
is much shorter than the  typical pulse length used in the 
Josephson junction qubits experiments and therefore the 
results from steady-state saturation theory can be used. 
However, in the unsaturated regime, 
$T^{\star}\approx T_1\approx 8 \times 10^{-8}$ sec at $T=0.1$ K, 
so transient effects can be important.

{\it High Frequency Charge Noise:}
At high frequencies (HF) ($\hbar\omega\gg k_BT$) the dielectric loss 
tangent is dominated by resonant (RES) absorption processes (i.e., 
$\tan\delta_{HF}\simeq\tan\delta_{RES}$) 
\cite{Arnold1976,Phillips,comment:REL_abs}:
\begin{equation}
\tan \delta_{HF}(\omega,T) = \frac{\pi p^{2} P_0}{3 \varepsilon'}
\tanh(\hbar\omega/2k_BT)\frac{1}{\sqrt{1+J/J_c(\omega,T)}},
\label{HFloss}
\end{equation}
where $J_c(\omega,T)=3\hbar^2\varepsilon^{\prime}v/(2p^2T_1 T_2)$ and 
$v$ is the speed of light in the solid. 
Eq.~(\ref{HFloss}) comes from integrating over the TLS distribution 
\cite{Arnold1976}. However, if no integration is done due to the
small number $N_0$ of TLS, e.g., in the tunnel junction barrier,
$\tan \delta_{HF}(\omega,T)=N_0 p^{2}T_2/(3\varepsilon^{\prime}\hbar \Omega) 
\times \mbox{tanh}(\hbar \omega/2k_BT)[1+J/J_c(\omega,T)]^{-1}$. So for
high intensities the $(J/J_c)^{-1/2}$ dependence of $\tan \delta$ 
becomes a $(J/J_c)^{-1}$ dependence. The frequency and temperature
dependence of $\tan\delta$, and hence of $S_{Q}(\omega)$, depends
on $J/J_c(\omega,T)$.

At low intensities ($J\ll J_c(\omega,T)$) in the unsaturated steady-state
resonant absorption regime, the dielectric loss tangent is constant:
\begin{equation}
\tan \delta_{HF} \simeq \pi p^{2} P_0/\left(3\varepsilon^{\prime}\right)
\label{IntrinsicLoss}
\end{equation}
For $\varepsilon_r=10$, $P_0 \approx 10^{45}$(Jm$^3$)$^{-1}$ 
\cite{ccyjjf,comment:P0}, and $p=3.7$ D (which corresponds to 
the dipole moment of an OH$^-$ impurity \cite{Golding1979}), we 
estimate $\tan\delta_{HF}\approx 1.8\times 10^{-3}$.
This result agrees well with the value of $\delta\simeq 1.6\times 10^{-3}$
reported in Ref.~\cite{Martinis2005}. In this regime the 
charge noise is constant:
$S_Q/e^2\simeq 4\pi\hbar C p^{2} P_0/(3e^2 \varepsilon^{\prime})$.
For $C=$ 7 aF, $S_Q/e^2\simeq 2\times10^{-16}$ Hz$^{-1}$.

For high field intensities $J\gg J_c(\omega,T)$, we obtain 
$\tan \delta_{HF}=\pi p^{2} P_0/(3 \varepsilon^{\prime})
\times\sqrt{J_c(\omega,T)/J}$. The $J^{-1/2}$ dependence of $\tan \delta$
has been found 
for materials such as amorphous SiO$_2$ and amorphous SiN$_x$ 
\cite{Martinis2005}. In the saturated resonant 
absorption regime the charge noise is given by 
\begin{equation}
\frac{S_Q(\omega,T)}{e^2}\simeq \frac{4\hbar C}{e^2}\frac{\pi p^{2} P_0}
{3\varepsilon^{\prime}}\sqrt{\frac{J_c(\omega,T)}{J} }. 
\label{eq:S_Q_HF_HI}
\end{equation}
So the frequency and temperature dependence of the noise is determined by
$T_1$ and $T_2$:
$S_Q(\omega,T)\sim \sqrt{J_c(\omega,T)}\sim \left(T_1 T_2\right)^{-1/2}$.
Experiments and theory find that $T_2^{-1}\sim T^m$ where $m$ ranges from
1 to 2.2 \cite{Black1977,Graebner1979,Bernard1979,Golding1982,Hegarty1983,
Schickfus1988,herve,Enss1996short}. $T_2$ decreases with increasing frequency
\cite{Graebner1979} but the exact frequency dependence is not known, so
we will ignore it in what follows.
$T_1$ in the symmetric case with $\hbar\omega=E=\Delta_o$ is given by 
Eq.~(\ref{eq:T1inv}): $T_1^{-1}\sim \omega^3$, implying that at high 
frequencies and intensities $S_Q(\omega,T)\sim \omega^{3/2}T^{m/2}$,
where $m/2$ varies between 0.5 and 1.1.
If there are only a very few TLS, then 
$S_Q(\omega,T)\sim J_c(\omega,T)\sim \left(T_1 T_2\right)\sim \omega^3 T^{m}$.
Although the experimental frequency dependence was reported to be linear
\cite{Astafiev2004}, the scatter in the data is large enough to 
allow for a steeper frequency dependence. In fact, the dependence is 
much steeper for data at the degeneracy point.

Experimental measurements \cite{Schickfus1977} on SiO$_2$ at $f=10$ GHz 
and 0.4 $< T < $1 K find $J_c(\omega,T)=25$ 
mW/cm$^2\times (T/0.4 $K$)^4$, and that
$J/J_c(\omega,T)$ varies between 
$10^{-2}$ and $10^{4}$. This implies that the charge noise 
$S_Q/e^2$ should vary between $2\times 10^{-16}$ Hz$^{-1}$ and 
$2\times 10^{-18}$ Hz$^{-1}$ at $T=0.1$ K. However, other measurements 
have found $J_c(\omega,T)\sim \omega^nT^2$ \cite{Arnold1976}, where $n$ 
is equal to either 0 \cite{Arnold1974,Bernard1979} or 2 
\cite{Graebner1979,Bachellerie1977}. If $J_c(\omega,T)\sim\omega^{2}$, 
then $\tan\delta(\omega)\sim\omega$, and $S_{Q}(\omega)\sim\omega$ 
at high frequencies and high intensities, which agrees with the recent 
experiments by Astafiev {\it et al.} \cite{Astafiev2004}.
It would be interesting measure the temperature dependence of $S_{Q}$
experimentally. 

While currently there are no direct experimental measurements of 
the HF charge noise, in Ref.~\cite{Astafiev2004} $S_Q(\omega)$ has 
been deduced by measuring the qubit relaxation rate $\Gamma_1$ versus the 
gate induced charge $q$ for a Cooper pair box with a 
capacitance $C_b$, Josephson energy $E_J$ (in the GHz range), and 
electrostatic energy $U=2eq/C_b$. From
$\Gamma_1=\pi S_U(\omega) \sin^2(\theta)/2\hbar^2$, 
where $\sin\theta=E_J^2/(E_J^2+U^2)$ and $\hbar\omega$ equals the qubit energy
splitting ($\hbar\omega=\left(U^2+E^2_J\right)^{1/2}$), and 
$S_U(\omega)=(2e/C_b)^2 S_q(\omega)$, they obtained the charge noise 
$S_q(\omega)$ at high frequency \cite{Astafiev2004}. We can compare our
results with these experiments by reversing this procedure 
to find $\Gamma_1(q)$ from $S_q(\omega)$. We find for saturated TLS that
$\Gamma_1$ at the maximum ($q=0$) is of order $10^{8}$ s$^{-1}$ 
\cite{comment:Gamma1} and increases
as $E_J^2$ in good quantitative agreement with the experimental results
from Fig.~2 and 3 of Astafiev {\it et al.} \cite{Astafiev2004}.
We find that $\Gamma_{1,max}$, the maximum value of $\Gamma_1$
(at $q=0$), increases with the frequency $f=E_J/h$ in the saturated
regime but is independent of frequency at low intensities $J\ll J_c(\omega,T)$.

{\it Low Frequency Charge Noise:}
We now show that we can recover the low frequency $1/f$ charge noise
using Eq.~(\ref{eq:S_Q_Tan_delta}). At low frequencies (LF) where
$\hbar\omega\ll k_B T$, 
only the relaxation absorption process contributes: 
$\tan\delta_{LF}\simeq\tan\delta_{REL} =
\pi p^2 P_0/(6\varepsilon')$ \cite{Classen1994}.
Eq.~(\ref{eq:S_Q_Tan_delta}) gives \cite{Faoro2006,Kogan96}
\begin{equation}
\frac{S_Q(f)}{e^2}=\frac{2k_BT}{e^2/2C}\tan \delta_{LF}\frac{1}{2\pi f}
=\frac{1}{3}\Omega P_0 k_BT \Bigr(\frac{p}{eL}\Bigr)^2\frac{1}{f}.
\label{eq:SQ_LF}
\end{equation}

To estimate the value of $S_Q$, we use $p=3.7$ D, 
$P_0 \approx 10^{45}$ (Jm$^3$)$^{-1}$, $L=400$ nm, and $A=40\times 800$ nm$^2$. 
At $T=100$ mK and $f=1$ Hz, we obtain $S_Q/e^2=2\times10^{-7}$ Hz$^{-1}$,
which is comparable to the experimental value of $4\times 10^{-6}$ Hz$^{-1}$
deduced from current noise \cite{Astafiev2006}.
As Eq.~(\ref{eq:SQ_LF}) shows, the standard TLS 
distribution gives low frequency $1/f$ charge noise that is 
linear in temperature, while experiments find a quadratic 
temperature dependence \cite{Kenyon2000,Astafiev2006}. 
This implies that 
at low frequencies and temperatures contributions from other mechanisms 
may dominate the charge noise \cite{Faoro2005}.

To conclude, we have shown that the frequency and temperature
dependence of high frequency charge noise in Josephson junction devices
depends on the ratio $J/J_c(\omega,T)$ of the electromagnetic
flux to the critical flux. Using the standard theory of two level
systems with a flat density of states, we find that the
charge noise at high frequencies can increase linearly with
frequency and temperature if $J/J_c(\omega,T)\gg 1$. This agrees 
with the conclusions of recent experiments 
on the high frequency charge 
noise in Josephson junction qubits \cite{Astafiev2004} which our 
estimates show are in the strongly saturated limit. 

This work was supported by ARDA through ARO Grant W911NF-04-1-0204,
and by DOE grant DE-FG02-04ER46107.
M. C. wishes to thank Fred Wellstood for useful discussions.


\begin{thebibliography}{36}
\bibitem[{com({\natexlab{a}})}]{comment:REL_abs}
\bibinfo{note}{The relaxation absorption loss at HF is given by $\tan
  \delta_{REL}\simeq \pi p^2 P_0/ (3\epsilon^{\prime})\times a (\pi k_B
  T)^3/(12\omega)$. We found that $\frac{\tan\delta_{REL}}
  {\tan\delta_{RES}}\simeq 5\times 10^{-6}$ for $J\ll J_c$ and $\simeq 2\times
  10^{-2}$ for $J\gg J_c$ at $0.1$ K and $10$ GHz.}
\end{thebibliography}
\end{document}